\documentclass[useAMS,usenatbib, usegraphicx]{mn2e}

\usepackage{epsfig}
\usepackage{amsmath}
\usepackage{amssymb}
\usepackage{natbib}
\newcommand{\kms}{\ensuremath{{\rm km\,s}^{-1}}}
\newcommand{\msun}{\ensuremath{M_{\odot}}}
\newcommand{\hills}{\citet{1988Natur.331..687H}}

\newcommand{\browna}{\citet{2005ApJ...622L..33B}}

\newcommand{\browns}{\citet{2006ApJ...640L..35B}}
\newcommand{\brownsp}{\citep{2006ApJ...640L..35B}}
\newcommand{\genzel}{\citet{2003ApJ...594..812G}}
\newcommand{\yutremaine}{\citet{2003ApJ...599.1129Y}}
\newcommand{\perets}{2006astro.ph..6443P}
\newcommand{\ghez}{2005ApJ...620..744G}
\newcommand{\infinity}{{\infty}}
\newcommand{\apj}{ApJ}
\newcommand{\apjl}{ApJ}
\newcommand{\mnras}{MNRAS}
\newcommand{\aj}{AJ}

\newcommand{\nat}{Nat}

\newcommand{\aap}{A\&A}
\newcommand{\letter}{{paper}}
\newcommand{\sag}{Sgr~A*}
\newcommand{\rmd}{{\rm d}}
\newcommand{\physrep}{Physics Reports}
\newcommand{\aaps}{A\&AS}
\title[Hypervelocity Stars from
the Galactic Centre]{Production of Hypervelocity Stars through Encounters with
Stellar-Mass Black Holes in the Galactic Centre}

\author[O'Leary \& Loeb]{Ryan M.\
O'Leary\thanks{E-mail:roleary@cfa.harvard.edu} and Abraham
Loeb\thanks{E-mail:aloeb@cfa.harvard.edu}\\ Harvard University, Department of Astronomy, 60 Garden St., Cambridge, MA 02138, USA\\}

\begin{document}
\maketitle

\begin{abstract}

  Stars within 0.1\,pc of the supermassive black hole \sag\ at the
  Galactic centre are expected to encounter a cluster of stellar-mass
  black holes (BHs) that have segregated to that region.  Some of
  these stars will scatter off an orbiting BH and be kicked out of the
  Galactic centre with velocities up to $\sim\,2000\,$\kms.  We
  calculate the resulting ejection rate of hypervelocity stars (HVSs)
  by this process under a variety of assumptions, and find it to be
  comparable to the tidal disruption rate of binary stars by \sag,
  first discussed by \hills.  Under some conditions, this novel
  process is sufficient to account for all of the hypervelocity
  B-stars observed in the halo, and may dominate the production rate
  of all HVSs with lifetimes much less than the relaxation time-scale
  at a distance $\sim 2\,$pc from \sag\ ($\gtrsim 2\,$Gyr).  Since
  HVSs are produced by at least two unavoidable processes, the
  statistics of HVSs could reveal bimodal velocity and mass
  distributions, and can constrain the distribution of BHs and stars
  in the innermost $0.1\,$pc around \sag.
\end{abstract}

\begin{keywords}
Galaxy:centre--Galaxy:kinematics and dynamics--stellar dynamics
\end{keywords}

\section{Introduction}
Hypervelocity stars (HVSs) have velocities so great ($\sim1000\,$\kms)
that they are gravitationally unbound to the Milky Way galaxy.  Since the
discovery of the first HVS by \browna, six additional HVSs have been found
\citep{2005Apj...634L.181E, 2005A&A...444L..61H, 2006ApJ...640L..35B,
2006ApJ...647..303B}.  The age and radial velocities of all but one HVS are
consistent with them originating from the Galactic centre\footnote{In the
case of \citet{2005Apj...634L.181E}, the star appears to originate from the
Large Magellanic Cloud.}, the most natural site for producing them (Hills
1988).  As more HVSs are found, they can be used to constrain many
properties of both the Galactic centre as well as the Milky-Way galaxy as a
whole, providing information about the Galactic potential
\citep{2005ApJ...634..344G,2006astro.ph..8159B}, the merger history of the
\sag\ \citep{2005astro.ph..8193L,2006astro.ph..7455B}, as well as the mass
distribution of stars near \sag.

Nearly 17 years before the initial discovery by \browna, \hills\ first
proposed that HVSs should populate the galaxy and provide indirect evidence
for the existence of a supermassive black hole (SMBH) in the Galactic
centre. \hills\ showed that when a tight stellar binary (with a separation
$<0.1\,$AU) gets sufficiently close to a SMBH, it will be disrupted by its
strong tidal field. Consequently, one member of the binary could be ejected
from the Galactic centre with sufficient energy to escape the gravitational
potential of the entire galaxy
\citep{1991AJ....102..704H,2005MNRAS.363..223G}.  The other binary member
is expected to remain in a highly eccentric orbit around the SMBH
\citep[see, e.g.,][]{2003ApJ...592..935G,2006MNRAS.368..221G}.
\yutremaine\ analysed the production rate of HVSs in more detail,
specifically for stars originating near the SMBH in our galaxy, \sag.  The
authors also corrected the initial calculation of \hills\ and accounted for
the diffusion of hard binaries into the ``loss--cone'', finding the
production rate of HVSs to be $\sim\,10^{-5}\,$yr$^{-1}$, nearly three
orders of magnitude below the previous estimate.  \yutremaine\ also looked
at two additional mechanisms for producing HVSs.  They found that the rate
could easily be higher ($\sim\,10^{-4}\,$yr$^{-1}$) if \sag\ had a massive
binary companion \citep[see
also][]{2005astro.ph..8193L,2006astro.ph..7455B,2006astro.ph..4299S}, and
also determined that star-star scattering near \sag\ resulted in a nearly
undetectable rate, due to physical collisions among the two stars.

In examining the rate estimate from \yutremaine\, one may adopt a simple
mass function (MF) for stars, $\rmd n/\rmd m \propto m^{-\beta}$ to
estimate the total number of B--type HVSs in the Galactic halo.  Using the
Salpeter MF slope $\beta = 2.35$ and a minimum mass limit of $0.5\,\msun$
one expects there to be about $\sim\,80$ HVSs with masses between $3$ and
$5\,\msun$, consistent with the observed rate for such stars \brownsp.
However this is an overestimate.  The Salpeter MF overestimates the total
number of stars with mass $>\,3\,\msun$, since star formation occurs
continuously over time near the Galactic centre and massive stars have
short lifetimes \citep{astern,\perets}. Accounting for continuous star formation
as well as a more realistic distribution of binary parameters,
\cite{\perets} found a lower total HVS rate of $5\times 10^{-7}\,$yr$^{-1}$
yielding $\sim 1$ HVS in the entire galaxy with mass between $3$ and
$5\,\msun$. The discrepancy between this calculation and the number of
observed HVSs may be overcome by massive perturbers in the Galactic centre,
which reduce the relaxation time and increase the rate to that observed
\citep{\perets}.  
Nevertheless, their results are highly sensitive to the mass
distribution of the perturbers.  In their calculations the authors
use a very flat slope for the distribution of masses which has not
been well constrained by observations.  In their estimate, they assume
that all stars are initially relaxed and therefore fill the entire
energy-momentum space.  The lifetimes of the stars observed by
\browns\ are very similar to the enhanced relaxation time, except in
the most optimistic case of \citet{\perets}.  We argue then that the
total number of B-type HVSs given by \cite{\perets} may still be an
overestimate, and an additional source of HVSs may be required.

In this \letter, we propose a novel source of HVSs in the Milky Way: in the
dense stellar cusp near \sag\, {\it stars scatter off stellar-mass black
holes} (BHs) that are segregated there \citep{1993ApJ...408..496M} and
recoil out of the Galactic centre. The existence of stars on radial orbits
originating from the Galactic centre was first proposed by
\cite{2000ApJ...545..847M} as evidence for a segregated BH cluster. Here,
we show that there should exist a high-velocity tail of stars, significant
enough in number to account for some if not all of the HVSs observed by
\browns.

This \letter\ is organised as follows.  In \S \ref{stars} we describe
the distribution of stars and BHs around \sag. We
describe in detail our assumptions and calculations in \S \ref{theory},
and present our results in \S \ref{results}.  Finally, in \S
\ref{discussion}, we discuss the observational implications and
prospects for future work.

\section{The cusp of stars and BHs around \sag}
\label{stars}
Using near-infrared adaptive optics imaging, \genzel\ showed that the
density of stars in the innermost $\sim 2\,$pc of the Galactic centre
is well-fit by a broken power-law radial profile, $\rho(R) = 1.2
\times 10^6 (R/0.4\,{\rm pc})^{-\alpha}\,\msun\,$pc$^{-3}$, where
$\alpha \approx 1.4$ for $R < 0.4\,$pc and $\alpha \approx 2.0$ for $R
> 0.4\,$pc, assuming that \sag\ is at a distance of $8\,$kpc
\citep{2005ApJ...620..744G,2005ApJ...628..246E}. Additional
observations of the background light density show that the cusp might
be slightly shallower in the inner few arcseconds
\citep{schoedel07}. Indeed, for a dynamically relaxed stellar system,
stars near a SMBH are expected to be in a cusp profile with $1.5 <
\alpha < 1.75$, depending on the mass distribution of the stars
(\citealp{1976ApJ...209..214B,1977ApJ...216..883B}; but see also
\citealp{2006ApJ...645L.133H}).  \genzel\ normalised the density
profile by assuming that the total mass of stars within $1.9\,$pc of
\sag\ is $\approx 3.2\times 10^6\,\msun$, and that the mass
distribution follows the same density profile as the observed stars
with a constant mass--to--light ratio. Invariably, stellar evolution
and mass segregation should alter the mass--to--light ratio at
different radii \citep{2006ApJ...645L.133H}, and cause the actual
density profile to deviate from the observed number counts of stars.
In addition to determining that the stars are in a cusp profile, the
observations also indicate that the stars have formed mostly
continuously with a standard initial MF
\citep{astern,2003ApJ...594..812G}.

Besides the observed cusp of stars, there are two additional
structures found near \sag\ that suggest ongoing star formation very
close to the SMBH.  Within 1 arcsec ($\approx\,0.04\,$pc) of \sag\ is
a nearly isotropic cluster of massive B--type stars, the so-called
``S--stars'' \citep{2005ApJ...628..246E,2005ApJ...620..744G}, whose
origin remains a mystery (see \citealp{2005PhR...419...65A} for a
review; see also \citealp{2006astro.ph..6443P} and references
therein.).  Outside this cluster, at 1--13 arcsec (0.04--0.5\,pc), is
at least one disk of young stars, with a stellar population distinctly
younger than the S--stars
\citep{2003ApJ...594..812G,2006ApJ...643.1011P}.  The spectral
identification of $\gtrsim 30$ post main sequence blue supergiants and
Wolf-Rayet stars constrains the age of the disk to $\lesssim 8\,$Myr
\citep{2006ApJ...643.1011P}.  The stars have top-heavy initial mass
function with $\beta \approx 1.35$
(\citealp{2006MNRAS.366.1410N,2006ApJ...643.1011P}; see also
\citealp{2005MNRAS.364L..23N}). Within the same observations of the disk,
\citet{2006ApJ...643.1011P} found the density of OB--type stars to
fall off more steeply than the observed cusp of stars, with no
positive detections outside of $0.5\,$pc.

Dispersed throughout the stars there should be a group of stellar-mass
BHs brought there through mass segregation
\citep{1993ApJ...408..496M,2000ApJ...545..847M}.  Assuming that all
the stars in the stellar cusp around \sag\ formed $10\,$Gyr ago,
\citet{2000ApJ...545..847M} predicted that there should be $\sim
2.5\times 10^4$ BHs within the central pc of the Milky Way.  In their
analysis, \citet{2000ApJ...545..847M} estimated that $1.6\%$ of the
stellar mass is in BHs given an approximately Salpeter MF.  All BHs
within $5\,$pc relax to the central pc through mass segregation, where
they form an $\alpha = 1.75- 2.0$ cusp.  More detailed studies support
this argument. \citet{2006ApJ...645L.133H} numerically solved the
time--dependant Boltzmann equations for a four mass model near \sag\
with $10\,\msun$ BHs, $1.4\,\msun$ neutron stars, $1\,\msun$ stars,
and $0.6\,\msun$ white dwarfs. They found that there should be
$\approx 1.8\times 10^3$ BHs and $\approx 3\times 10^4\,\msun$ of
stars within $0.1\,$pc of \sag\ in an $\alpha_{\rm BH} \approx 2.0$ ,
$\alpha_{*} \approx 1.4$ cusp respectively. In addition, large $N$
Monte-Carlo simulations of relaxation and stellar evolution in the
Galactic centre found the formation of a similar cusp of BHs
\citep{2006astro.ph..3280F}.  However, in these simulations the
density of BHs at $0.1\,$pc was found to be ten times higher and the
density of stars to be about five times lower than in
\citet{2006ApJ...645L.133H}.  Very few observations constrain number
of stellar-mass BHs considered here, however X-ray observations seem
to place an upper limit of $\sim 40,000$ BHs in the inner parsec
\citep{deegan}.

\section{Theory and assumptions}
\label{theory}

To calculate the rate at which stars are ejected from the Galactic
centre through their scattering off BHs there, we generalise the
analysis of \yutremaine\ to multi--mass systems \citep[in a form
similar to that of][]{1969A&A.....2..151H}.  In our calculations we
assume that \sag\ has a mass $M_{\rm smbh} = 3.5\times 10^6\,\msun$
and is at a distance of $8\,$kpc \citep{2005ApJ...620..744G,2005ApJ...628..246E}.

\subsection{Two--body scattering}
We would like to identify the conditions under which a scattering between
a BH and a star would result in the ejection of a HVS. In each scattering,
a star of mass $m$ and velocity $\bmath v$ undergoes a change in velocity
(\citealp{1987gady.book.....B}, Eqs. 7-10a - 10b)
\begin{eqnarray}
\label{deltahv}
{\bmath \delta \hat{\bmath{v}}} = \frac{2m'bw^3}{G(m'+m)^2} \left(1+\frac{b^2w^4}{G^2(m'+m)^2}\right)^{-1} \hat{\bmath{w}}_{\perp} \nonumber\\
- \frac{2m'w}{m'+m} \left(1+\frac{b^2w^4}{G^2(m'+m)^2}\right)^{-1} \hat{\bmath{w}},
\end{eqnarray}
where $b$ is the impact parameter of the encounter, $w = |{\bmath w}|
= |{\bmath v}-{\bmath v}'|$ is the relative velocity at infinity, $m'$
is the mass of the BH, and ${\bmath v}'$ is the BH's velocity. The
direction of ${\bmath \delta \hat{v}}$ is determined by $\hat{\bmath
  w}$ and $\hat{\bmath w}_{\perp}$, the unit vectors along and
perpendicular to $\bmath{w}$, respectively.  

In some instances, the BH may tidally distort (or even disrupt) the
incoming star, causing the BH-star system to lose energy. 
At closest approach, the stars will have a relative velocity of
$\sqrt{w^2 + 2 G (m+m')/b_{\rm min}}$, where $b_{\rm min}$ is the
distance of closest approach between the star and BH.  In order to be
conservative, we assume that the star is composed of two halves, each
with mass $m/2$, separated by a distance of two stellar radii, $2
R_*$. This is chosen in order to have the energy loss diverge when
$b_{\rm min}=R_*$.  Then, in the impulse approximation, the total
amount of energy lost is approximately
\begin{equation}
\label{deltae}
\Delta E \approx \frac{2 G^2 m m'^2 R_*^2}{(w^2 + 2 G (m + m') / b_{\rm min})(b_{\rm min}^2-R_*^2)^2}.
\end{equation}
Thus, the final relative velocity between the star and BH is reduced to,
\begin{equation}
\label{finalrelvel}
w_{\rm f} = \sqrt{w^2 - 2 \Delta E/m}.
\end{equation}
If we now assume that the final relative velocity is in the same
direction as in Equation~(\ref{deltahv}), we get the final change in
velocity,
\begin{eqnarray}
\label{deltav}
{\bmath \delta \bmath{v}} = \frac{2m'b w_f w^2}{G(m'+m)^2} \left(1+\frac{b^2w^4}{G^2(m'+m)^2}\right)^{-1} \hat{\bmath{w}}_{\perp} \nonumber\\
- \frac{m'}{m'+m} \left(w - w_f + 2w_f\left(1+\frac{b^2w^4}{G^2(m'+m)^2}\right)^{-1}\right) \hat{\bmath{w}}.
\end{eqnarray}
Equation~(\ref{deltav}) correctly reduces to Equation~(\ref{deltahv}) as $w_f$
approaches $w$.

We are seeking circumstances under which the star's velocity
at infinity is greater than some threshold ejection speed $v_{\rm ej}$,
\begin{equation}
\label{vinf}
v_{\infinity}^2 = |\bmath{v}+\bmath{\delta v}|^2 - \frac{2 G M_{\rm
      smbh}}{r} \geq v_{\rm ej}^2,
\end{equation}
where $r$ is the distance from \sag\ at which the encounter
occurred. For the distance of $55\,$kpc, at which many HVSs have been
observed, the star would have a velocity $v_{\rm 55} \approx
[v_\infinity^2-(800\,\kms)^2]^{1/2}$ \citep{1987AJ.....94..666C}. 

For our calculations we determine $R_{*}$ by fitting a broken
power-law to the solar metallicity stellar models of
\citet{1992A&AS...96..269S}, and do not consider ejections where the
star is tidally disrupted by the BH (i.e. $\Delta E \gtrsim
Gm^2/R_*$). 
For a typical star of $1\,\msun$ at a distance of $.01\,$pc, this
condition would give a minimum closest approach of $b_{\rm min}
\approx 2.07 R_*$.  At $.001\,$pc, the star and BH could get as close
as $1.57 R_*$ before tidal disruption.

\subsection{Total rate}
The population of stars near \sag\ can be described by the
seven--dimensional distribution function (DF) $f_*({\bmath r}, {\bmath
v}, m)$, where $m$ is the star's mass, ${\bmath v}$ is the star's
velocity, and ${\bmath r}$ is the star's position relative to \sag\
which is located at ${\bmath r} = \bmath{0}$. Similarly, we describe
the DF of BHs by $f_{\rm BH}({\bmath r}', {\bmath v}', m')$.

In our analysis we restrict our attention to stars scattering off BHs only,
since in star--star scattering, physical collisions between the stars limit
the total production rate of HVSs (Yu \& Tremaine 2003).   
Thus, the probability for a test star to encounter a BH at an impact parameter
$b$ within an infinitesimal time interval $\rmd t$ is
\citep{1960AnAp...23..467H,2003ApJ...599.1129Y}
\begin{equation}
\Gamma({\bmath r}, {\bmath v}) \rmd t= \rmd t \int b\,\rmd b \int w\, \rmd^3{\bmath v}' \int \rmd{\Psi} \int \rmd m'\, f_{\rm BH}({\bmath r}, {\bmath v}', m'),
\end{equation}
where $m'$ is the BH mass, and $\Psi$ is the angle between the $({\bmath
  v}, {\bmath v}')$-plane and $({\bmath v} - {\bmath v'}, {\bmath
  \delta v})$-plane.  The integrated rate at which all stars 
  undergo such encounters in a small volume, $\rmd^3{\bmath r}$,
  around position ${\bmath r}$ is
\begin{equation}
\mathcal{R}({\bmath r}) = \int \rmd^3 {\bmath v}\int \rmd m\,f_*(\bmath r, \bmath v, m)\,\Gamma({\bmath r}, {\bmath v}).
\end{equation}
Finally, the total rate of ejecting stars with velocity $\geq v_{\rm ej}$
is
\begin{equation}
\label{finalrate}
\frac{\rmd N_{\rm ej}}{\rmd t} = \int \rmd^3{\bmath r} \mathcal{R}({\bmath r}),
\end{equation}
from some inner value $r_{\rm min}$ (which depends on the distribution
function of the stars, see, \S \ref{disfunct}) to $0.1\,$pc, where
we limit the integration over $b$, $w$, and $\Psi$ such that
$v_{\infinity} > v_{\rm ej}$, as in Eq.~(\ref{vinf}), and exclude all
encounters that tidally disrupt the star ($\Delta E \gtrsim
Gm^2/R_*$).

In order to find the total rate of HVSs we integrate
Eq.~(\ref{finalrate}) using a multidimensional Monte-Carlo
numerical integrator.  Because the integrand of
Eq.~\ref{finalrate} is not continuous, other integration routines
are unsuitable for our calculations. We integrate all of our results
until they converge to a residual statistical error of $< 10\,\%$.

\begin{table*}
  \caption{\label{table1} Stellar DF Models and Results.  The first four columns describe the different stellar distribution functions used in our analysis. 
    $\beta$ is the slope of the MF, which is normalised by the maximum mass (column 3) and minimum mass (column 4) 
    of the stars, as well as the total number of stars, as described in  \S~\ref{disfunct}. The remaining columns list the total integrated  rate of HVSs ($v_\infinity > 1000\,\kms$) for each model. From left to right the rates correspond to $m' = 10\,\msun$, $m' = 15\,\msun$ both 
    with $r_{\rm min} =  0.001\,$pc, and finally $m' = 15\,\msun$ with $r_{\rm min} = 0$.  Our results are described in more detail 
    in \S~\ref{results}.}
 
   \begin{tabular}{@{}lrrrrrr@{}}
  \hline
  Model     & $\beta$ & Maximum Mass & Minimum Mass  &      &              HVS Rate     &       \\
  &         &              &              & $m' = 10\,\msun$ &              $m' = 15\,\msun$    &  $r_{\rm min} = 0$  \\
             &         &     ($\msun$) &   ($\msun$)  & (yr$^{-1})$      &            (yr$^{-1}$)  & (yr$^{-1}$)    \\
\hline
  1a         &   4.85  &    20       &    3        &   $3.5 \times 10^{-9}   $ &   $1.5 \times 10^{-8}   $ &  $8.2\times 10^{-8}$   \\
  1b         &   3.85  &    20       &    3        &   $2.0 \times 10^{-9}  $ &   $7.8 \times 10^{-9}   $ &  $3.9\times 10^{-8}$   \\
  2a         &   2.35  &     3       &    0.5      &   $3.3 \times 10^{-7}   $ &   $9.8 \times 10^{-7}   $ &  $3.6\times 10^{-6}$   \\
  2b         &   2.35, 4.85  &   3   &    0.5      &   $4.4 \times 10^{-7}   $ &   $1.3 \times 10^{-6}   $ &  $4.8\times 10^{-6}$   \\
  \hline
\end{tabular}
\end{table*}

\subsection{Distribution functions}
\label{disfunct}
Unfortunately, there are no observations to constrain the expected
number or distribution of BHs near \sag\ \citep[see also,][]{deegan}.
For simplicity, we assume that all BHs have a mass of $10\,\msun$, are
distributed isotropically, and follow an $\alpha_{\rm BH}$ cusp
density profile so that
\begin{equation}
\label{bhdist}
f_{\rm BH} \propto E^{\alpha_{\rm BH} - 1.5} \delta(m' - 10\,\msun),
\end{equation}
where $E= (G M_{\rm smbh} / r - v'^2/2)$ is the negative specific
energy of a BH and $\alpha_{\rm BH} = 2$.  
Equation~(\ref{bhdist}) is normalised by integrating
Equation~\ref{bhdist} over $|v'| < \sqrt{2 G M/r}$ and $r$ so that there
are $N_{\rm BHs} = 1800$ BHs within $0.1\,$pc of \sag, consistent with
the calculations of \cite{2000ApJ...545..847M} and
\cite{2006ApJ...645L.133H}.  In addition, we look at a system of
$N_{\rm BHs} = 1800$ more massive BHs, $m' = 15\,\msun$, in order to
determine the importance of the BH MF on the rate of HVSs.  Because
the density power-law assumed here $\alpha_{\rm BH} = 2$ is steeper
than the $\approx 1.5$ slope derived for a ``drain limited''
population of BHs \citep[see Eq.~1 of][]{2004ApJ...606L..21A}, the
innermost regions of the integration (interior to $\lesssim 10^{-4} -
.001\,$pc) will have a number density that exceeds this theoretical
limit. Therefore, at the end of \S~\ref{results} we also look at an
approximately drain limited sample of BHs.

The stellar DF, on the other hand, should be much shallower than the
BH's within $\sim\,0.1\,$pc of \sag.  Observations as well as the
calculations of \cite{2006ApJ...645L.133H} suggest $\alpha_{*}
= 1.4$ in this region \citep{2003ApJ...594..812G}. This gives
\begin{equation}
\label{stardist}
f_{*} \propto E^{-0.1}  \frac{\rmd n}{\rmd m} ,
\end{equation}
where $\rmd n / \rmd m \propto m^{-\beta}$ is the current MF of stars.
We truncate Eq.~(\ref{stardist}) so that $f_* = 0$ when
\begin{equation}
\label{constraint1}
E > G (m' + m)/R_{*}.
\end{equation}
Stars would need to undergo physical collisions with the BHs, which
are rare and disruptive, in order to relax to a higher specific energy
\citep{1976ApJ...209..214B}. 
For a typical $1\,\msun$ star, this corresponds to an orbit with
minimum semimajor axis $\approx .007\,$pc, remarkably close to the
semimajor axis of S2 (S0-2) of $.005\,$pc
\citep{2005ApJ...620..744G,2005ApJ...628..246E}. Because we assume
that there are no stars with a tighter orbit, interior to the region
the number density of stars is much shallower, and follows an $\alpha_*
\approx .5$ power-law. In addition to the above constraint, we
explicitly exclude all orbits which would tidally disrupt the star,
\begin{equation}
\label{constraint2}
 r_{\rm closest} < R_{*} \left(\frac{m}{M_{\rm smbh}}\right)^{1/3},
\end{equation}  
where $r_{\rm closest}$ is the star's closest approach to the SMBH.
Although Equations~\ref{constraint1}~\&~\ref{constraint2} should
completely determine the innermost radius of the stars, $r_{\rm min}$,
we also set $f_{*} = 0$ for $r< r_{\min}$.  Except where explicitly
mentioned in our results, we conservatively set $r_{\rm min} =
0.001\,$pc. This is the innermost resolved radius of the simulations
of \citet{2006astro.ph..3280F}, where the authors found no significant
change in the DF caused by stellar
collisions.  

For the stellar DF, we use a total of four models, varying both $\beta$
and the normalisation. We do this both to constrain the total number
of HVSs as well test how sensitive the rate is on the model. For
clarity of our results, as well as to satisfy different observational
constraints, we separate the stars into two distinct populations.

For stars with mass $m > 3\,\msun$, we are constrained by the total
number of observed S--stars, as well as the star formation history in
the region.  There are $O(10)$ S--stars with mass $m \gtrsim 5\,\msun$
within $0.01\,$pc of \sag\
\citep{2005ApJ...620..744G,2005ApJ...628..246E}. We assume that this
reflects approximately a steady state and normalise our models (1a-b)
to 10 such stars subject to the above constraints (i.e.,
Eqs.~\ref{constraint1}~\&~\ref{constraint2}).  To have a complete
description we also need to know the slope of the current MF of stars,
$\beta$. Unfortunately, this is unconstrained for stars with $m
\lesssim 5\,\msun$.  However observations suggest that all of the
stars in the region have formed mostly continuously in time
\citep{astern,2003ApJ...594..812G}.  For completeness, we look at
two different values of $\beta$.  If the stars formed with a
standard Salpeter MF continuously throughout time, we would expect
${\rmd m}/{\rmd n} \propto m^{-2.35} \times t_m$, where $t_m$ is the
lifetime of the stars.  For $m \lesssim 10\,\msun$, $t_m \propto
m^{-2.5}$, therefore for Model 1a, we set $\beta = 2.35 + 2.5 = 4.85$.
In Model 1b, we set $\beta = 1.35 + 2.5 = 3.85$, consistent with a
flatter MF as observed in the young disks of massive stars
\citep{2006MNRAS.366.1410N,2006ApJ...643.1011P}. 
We have confirmed that the total amount of background light contributed
by the unresolved stars is very near, but does not exceed the best
estimates from the observed background light in the inner cusp
\citep{schoedel07}.

For stars with mass $0.5\,\msun < m < 3\,\msun$, we use two different
models.  For Model 2a, we use a Salpeter MF with $\beta = 2.35$. For Model
2b, we use a broken-power law with $\beta = 2.35$ for $0.5\,\msun < m
< 1\,\msun$, and $\beta = 4.85$ for $1\,\msun < m < 3\,\msun$, roughly
consistent with a population of stars that formed continuously over the
last 10$\,$Gyr.  In both cases, we normalise Eq.~(\ref{stardist})
by requiring that the total mass in stars for $r < 0.1\,$pc be equal
to $6\times10^4\,\msun$ \citep{2003ApJ...594..812G}.

\section{Results}
\label{results}

\begin{figure}
\begin{center}
\includegraphics[width=\columnwidth]{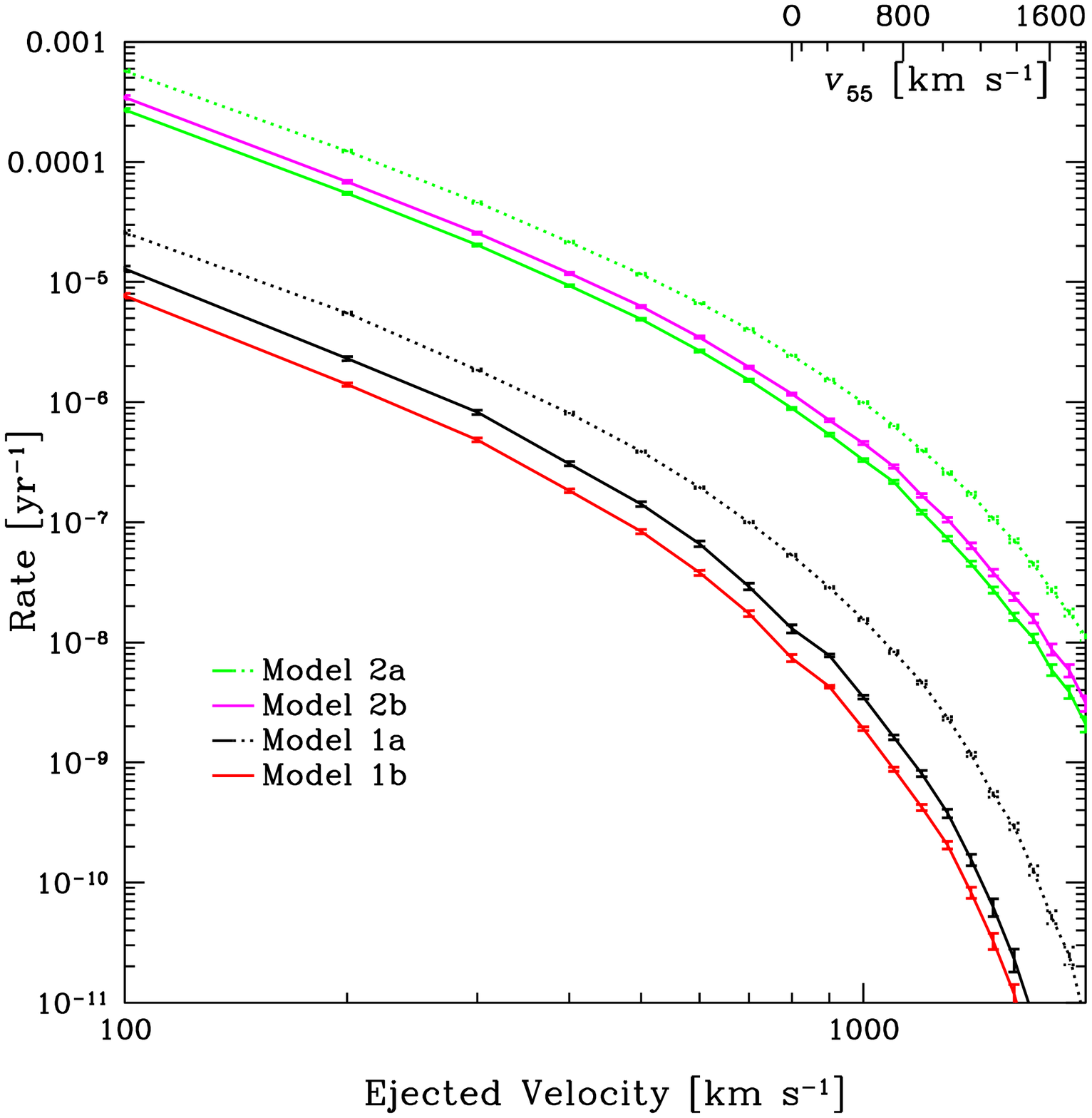}
\end{center}
\caption{\label{velocity} The total rate of stars with ejected
  velocities $v_{\infinity} > v_{\rm ej}$ versus $v_{\rm ej}$ for
  $r_{\rm min} = 0.001\,$pc. The coloured curves from top to bottom
  are for Models 2a (green), 2b (magenta), 1a (black) and
  1b (red).  The dashed curves are for the same stellar MFs but for a
  system of $15\,\msun$ BHs instead of $10\,\msun$.  The top axis is
  labelled to show the corresponding velocity at a galactocentric
  radius of $55\,$kpc, $v_{\rm 55}$.}
\end{figure}

In our results, we consider a star to be a HVS when $v_{\infinity} >
1000\,\kms$ ($v_{\rm ej} = 1000\,\kms$).  We list our results along
with a summary of each Model's MF in Table~\ref{table1}.  In
Figure~\ref{velocity}, we show how the total ejection rate depends on
the minimum ejection velocity, $v_{\rm ej}$, with $r_{\rm min} =
0.001\,$pc. For small ejection velocities ($v_{\rm ej} \sim
100\,\kms$) the rate decreases roughly as a power-law $\propto v_{\rm
  ej}^{-2.5}$ independent of the stellar MF or mass of the BHs.
However, for velocities $v_{\rm ej} \gtrsim 800\,\kms$ tidal dissipation as
well as physical collisions between stars and BHs begin to suppress
the rate of HVSs. In our Model 1a, with $m_{\rm min} = 3\,\msun$ and
$10\,\msun$ BHs, the ejection rate decreases rapidly from $3.5\times
10^{-9}\,$yr$^{-1}$ for $v_{\rm ej} = 1000\,\kms$ (with an observed
velocity of $v_{\rm 55} > 600\,\kms$ at 55\,kpc from \sag) down to
$6.3\times 10^{-11}\,$yr$^{-1}$ for $v_{\rm ej} = 1500\,\kms$ ($v_{\rm
  55} \gtrsim 1300\,\kms)$.  The velocity distribution of lower mass
HVSs ($\sim\,.5\,\msun$) shows a slightly weaker change in the rate of
HVSs in the high velocity range, as can be seen in
Figure~\ref{velocity}.  Thus, the observed distribution of HVSs
produced from this mechanism should be truncated sharply around
$\sim\,2000\,{\rm km~s^{-1}}$, in contrast to tidally disrupted
binaries which easily eject HVSs with velocities exceeding
$4000\,\kms$ \citep{2006MNRAS.368..221G,2006astro.ph..8159B}.

\begin{figure}
\begin{center}
\includegraphics[width=\columnwidth]{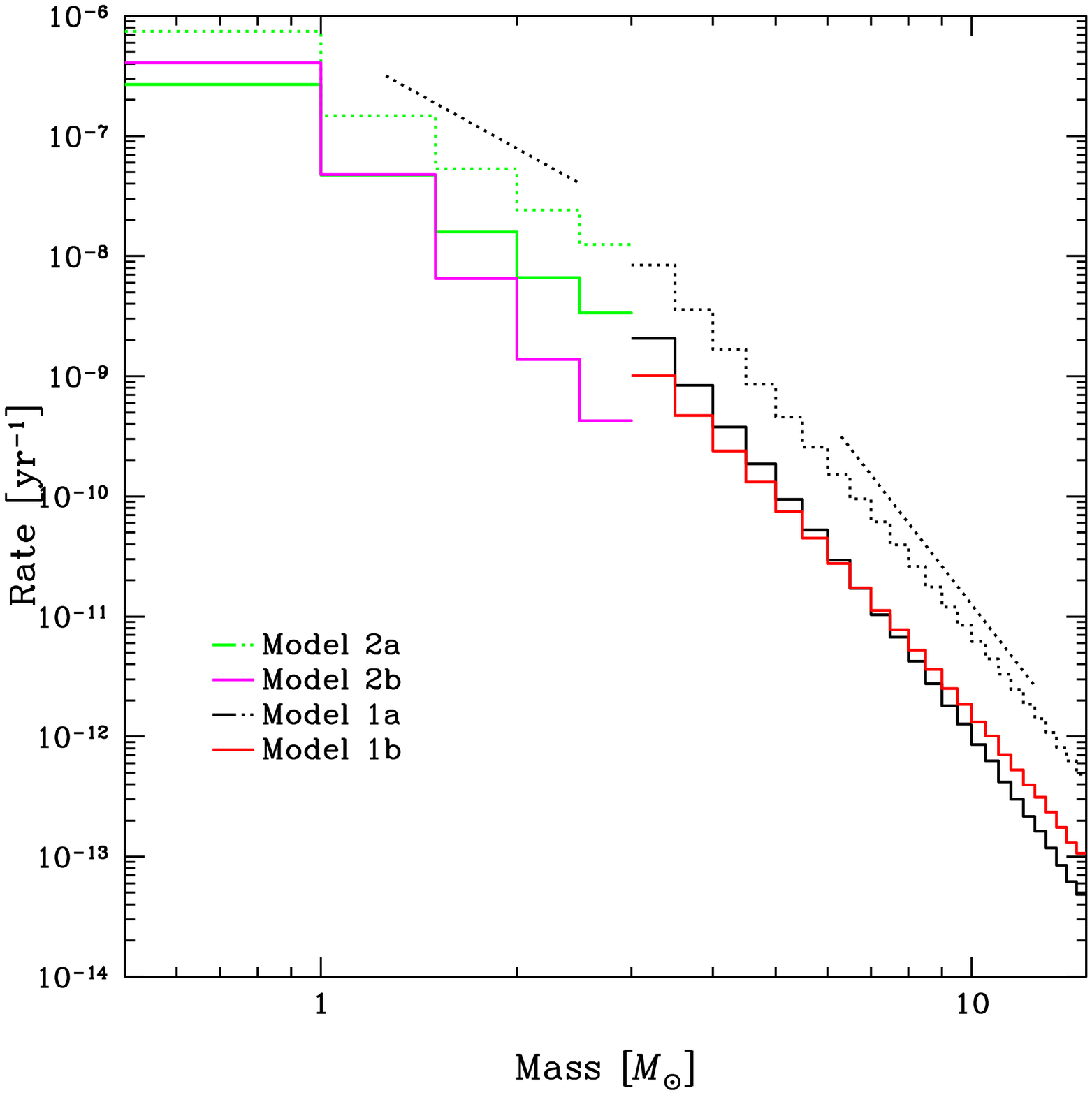}
\end{center}
\caption{\label{massdep} The total rate of HVSs ($v_{\rm ej} =
  1000\,\kms$,) in $0.5\,\msun$ intervals for $r_{\rm min} =
  0.001\,$pc.  The colours correspond to the same models as in
  Figure~\ref{velocity}. The top and bottom black `dotted' lines are
  for a power-law with slope -3, and -7 respectively for comparison.
  The discontinuity of the HVS rate at $3\,\msun$ is a result of our
  two different normalisations for Models 1 and 2.}
\end{figure}

Figure~\ref{massdep} shows the dependence of the rate of HVSs on the
stars' masses.  The rate of HVSs is plotted in
$0.5\,\msun$ intervals, so that the total rate of a model is the sum
over all mass bins.  For a cluster of single mass BHs, the mass
distribution of HVSs produced by this mechanism is steeper than the
stellar MF by $\sim 1$ for the low mass stars, and $\sim 2$ for higher
mass stars.  A steeper slope is expected since more massive stars will
receive a much smaller kick than less massive stars ($\sim \propto m'
/ (m + m')$). However, it remains unclear if these slopes will hold
for cluster of BHs with varied masses, where a few of the most massive
BHs may eject the majority of HVSs.  The steep slope of the HVS MF
makes it clear that very massive HVSs ($m \sim 10\,\msun$) can not
originate from this mechanism, unless there were significantly more stars
in the past.  There is also a clear discontinuity in the rate of
HVSs at $3\,\msun$. This just reflects the clear overdensity of high
mass stars in this region.


\begin{figure}
\begin{center}
\includegraphics[width=\columnwidth]{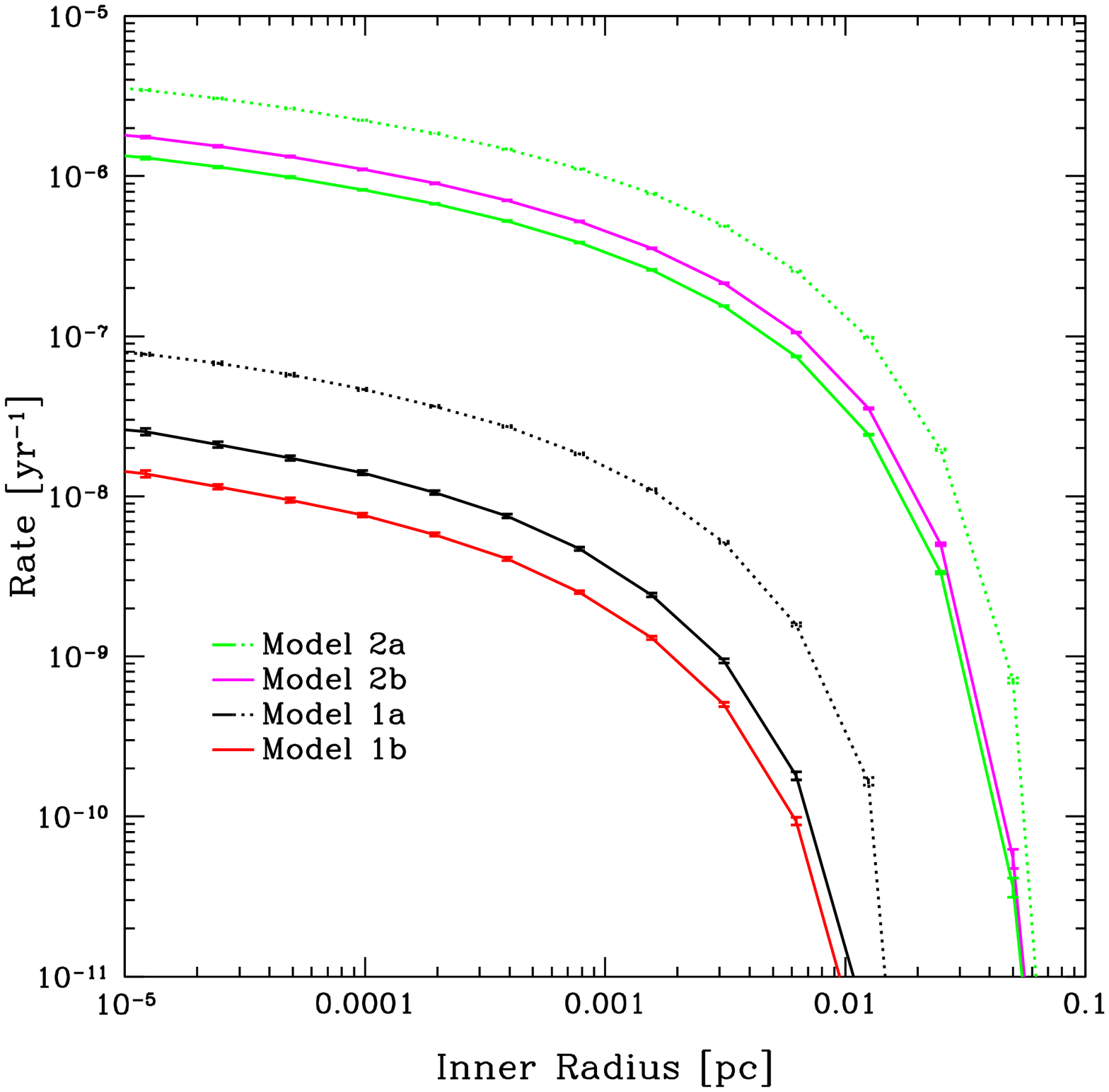}
\end{center}
\caption{\label{minrad} The HVS ejection rate ($v_{\rm ej} =
1000\,\kms$) versus the minimum radius of integration $r_{\rm
min}$. The lines are coloured as in Fig.~\ref{velocity}. Interior
to $0.001\,$pc, stellar collisions may flatten the cusp of low mass
stars \citep{1991ApJ...370...60M,2006astro.ph..3280F} and cause the rate of HVSs to
increase more slowly than shown in Figure~\ref{minrad}. For the high
mass stars in the region (Models 1a-b), the stars are less likely to
have physical collisions in their lifetimes, and may populate the
entire phase space allowed by Eqs.~(\ref{constraint1}) \&
(\ref{constraint2}). }
\end{figure}

The ejection rate also depends very strongly on the extent to which
stars populate the innermost regions of the cusp, as shown in
Figure~\ref{minrad}.  This is despite the constraints on the stellar
DF (c.f. Eqs.~\ref{constraint1} \& \ref{constraint2}), which limit
both the semimajor axis of stars to $\sim\,0.01\,$pc and the maximum
eccentricity of the orbit.  For $r_{\rm min} \lesssim 0.001\,$pc the
total rate increases with a shallow power-law index versus inner
radius, doubling its value by $10^{-4}\,$pc.  We caution, however,
that interior to $0.001\,$pc, stellar collisions may flatten the cusp
of low mass stars \citep{1991ApJ...370...60M, 2006astro.ph..3280F} and
cause the rate of HVSs to increase more slowly than shown in
Figure~\ref{minrad}. For the high mass stars in the region (Models
1a-b), the stars are less likely to have physical collisions in their
lifetimes, and may populate the entire phase space allowed by
Eqs.~(\ref{constraint1}) \& (\ref{constraint2}).

Since \browns\ initiated a targeted search for HVSs in the Galactic
halo, it is useful to forecast the number of observable HVSs that
would originate from encounters with stellar-mass BHs.  Current
observations indicate that there are $\sim 33 \pm 17$ hypervelocity
B-stars with masses between $3$--$5\,\msun$ (W. Brown, private
communication).  {\it Can scattered stars alone explain the abundance
  of the observed HVSs?}  The answer is positive but not under the
most conservative conditions assumed thus far, which predict only
$\sim 1 (N_{\rm BH}/1800)$ such HVSs, for $r_{\rm min} =0.001\,$pc.
If the masses of the BHs in the cluster were more typically $\sim
15\,\msun$, as might be expected from mass segregation as well as an
old population of BHs \citep{2004ApJ...611.1068B}, the number HVSs in
our Galaxy would be four times larger. This is still lower than the
number of observed HVSs. However, if we assume the most massive stars
populate the entire phase space within the constraints of
Eqs.~(\ref{constraint1}) \& (\ref{constraint2}), we expect there to be
$\sim 30$ HVSs in our Galaxy. If we relax our definition of a HVS and
set $v_{\rm ej} = 900\,\kms$, the ejection velocity of HVS7
\citep{1987AJ.....94..666C,2006ApJ...647..303B}, we get $\sim 50$
HVSs.
The assumption that the stars fill the entire phase space is in best
agreement with the observations of the S-stars, which appear to be a
relaxed system \citep{2005ApJ...620..744G}.  By integrating the DF in
this region, subject to the above constraints, we find that there
should be $\lesssim 0.3$ stars interior to 0.001$\,$pc.  In addition,
Equation~(\ref{constraint1}) limits the semimajor axis of any star to
$\sim 0.01$pc, close to the semimajor axis of S2 (S0-2), $\approx
0.005\,$pc.  Thus, the stars that are ejected from this region are
ones on eccentric orbits which bring them into this region, similar to
S14 (S0-16), which has a closest approach of $\approx 0.0002\,$pc
\citep{2005ApJ...620..744G}.  This is also consistent with the
resonant relaxation timescales in the region \citep[][however, note
our caution in
\S~\ref{discussion}]{1996NewA....1..149R,2006ApJ...645.1152H}. We also
note, however, that we are still being conservative in our
calculations.  We have excluded all collisional encounters, which at
the highest relative velocities may not actually disrupt the star or
have significant tidal energy dissipation. If stars and BHs do
populate the inner $10^{-4}\,$pc, any collisions would be at
velocities far greater than the escape velocity of the star. Our
calculations suggest that such encounters may contribute a factor of a
few more HVSs, and may offset any depletion of stars and BHs in the
region.  Since any such collisions would be brief and not very
luminous (unless coalescence follows at low impact speeds), it would
be difficult to identify star-BH collision events in external Galactic
nuclei.

Most recently, \citet{2007astro.ph..1600B} have detected a significant
number of marginally bound HVSs in the galactic halo.  They found that
these stars outnumber the unbound population by a factor of $\sim 2$.
Such a ratio of bound to unbound HVSs is consistent with the scenario
presented here, as well as the tidally ejected binary scenario \hills\
to within the statistical uncertainties
\citep{2007astro.ph..1600B}. If we look at the ratio of the rates of
stars ejected with velocities between $845\,\kms$ ($v_{55} =
275\,\kms$) and $920\,\kms$ ($v_{55} = 450\,\kms$), and stars ejected
with higher velocities, we find that there should be $\approx 2.6$
times more bound HVSs than unbound HVSs in a volume limited survey.
However, this analysis assumes that all such HVSs would be observable,
and does not account for any selection effects of the survey by
\citet{2007astro.ph..1600B}. With more HVS discoveries, a more
detailed statistical analysis may be able to determine the origin of
the HVSs.

As we pointed out in \S~\ref{disfunct}, the density profile of BHs
assumed thus far will exceed the ``drain limit''
\citep{2004ApJ...606L..21A} in the innermost regions of integration.
We have therefore approximated the drain limit for both $10\,\msun$ and
$15\,\msun$ BHs by using an $\alpha_{\rm BH} = 1.7$ power-law, which
we found to be the best fit to the Equation~1 of
\citet{2004ApJ...606L..21A} between $0.001$ and $0.1\,$pc.  We assumed
that, inside of $0.1\,$pc, there are $N_{\rm BH} \approx 12000$
$10\,\msun$ BHs and $\approx 5500$ $15\,\msun$ BHs.  We have found
that this increases the total rates calculated above by a factor of
$\sim 4$ for the $10\,\msun$ BHs and $\sim 2$ for the $15\,\msun$ BHs.
This is true, even if the BHs are drain limited only out to $\sim
0.01\,$pc.  Thus, for a drain limited population of $10\,\msun$ BHs
(interior to $0.01\,$pc), there would be $\sim 40$ hypervelocity
B-stars produced from BH-star scattering, with $r_{\rm min} = 0$.

\section{Summary and discussion}
\label{discussion}
We constructed a variety of models for the distribution of stars and
BHs in the innermost $0.1\,$pc of the Galactic centre.  Assuming that
the cusp and disk of stars we observe today represent a steady state
over the past $\sim\,100\,$Myr, we calculated the rate at which stars
will scatter off BHs and populate the Milky Way halo. We showed that
the total ejection rate of HVSs can be comparable to that produced by
the tidal disruption of binaries, and may exceed it for intermediate
mass stars. Our results are consistent with the total number of
ejected stars from Fokker-Plank simulations by
\citet{2006astro.ph..3280F}. 
Assuming that the B-stars are fully relaxed as is observed with the
younger S-Stars \citep{\ghez}, we can account for most, if not all, of
the stars observed by \browns.  We demonstrated that the velocity
distribution as well as the mass distribution of HVSs should be
truncated at high values for the BH scattering process compared to
binary disruption events (see Figs. 1 and 2).  Better statistics of
HVS detections could therefore determine the relative significance of
these two plausible channels.

In our study, the ejected stars originate from the inner $\sim 0.1\,$pc
near \sag, whereas the tidally disrupted binaries that produce HVSs
originally come from $\gtrsim\,2\,$pc
\citep{2003ApJ...599.1129Y,2006astro.ph..6443P}. The observations of the
young disk of stars \citep{2006ApJ...643.1011P}, as well as the cluster of
S-stars, suggest that the population of stars at 0.1\,pc from \sag\ should
be younger and more massive than at $2\,$pc.  This is further supported by
the steep drop in B-stars outside of 0.5\,pc
\citep{2006ApJ...643.1011P}. In addition, binaries with B-stars ($t_{\rm
ms} \lesssim 300\,$Myr) may also not be fully relaxed as has been assumed
in many previous calculations, and therefore the diffusion rate of the most
massive binaries into the loss-cone may be lower by orders of magnitudes.

In the survey of \citet{2006ApJ...647..303B}, the observed HVSs have
the same colour as blue horizontal branch stars (BHBs) and have yet to
be distinguished from B-stars with high resolution spectroscopy.  The
calculations shown here suggest that low mass stars, such as BHBs and
their progenitors, can become HVSs with high efficiency if they are in
the inner 0.1\,pc of \sag. 
Although tidal disruption rates may be enhanced by massive perturbers,
BHBs must undergo significant mass loss, causing them to widen or even
disrupt, and therefore are unlikely to be found in the tight binaries
which are disrupted \citep{2006astro.ph..6443P}. The spectroscopic
identification of one hypervelocity BHB star would be strong evidence
in support of the mechanism presented here, or an inspiralling massive
BH
\citep{2003ApJ...599.1129Y,2005astro.ph..8193L,2006astro.ph..7455B}. However,
the expected rate of hypervelocity BHBs depends on uncertain details
of stellar evolution and goes beyond the scope of this work.  A
population of BHBs ejected through this mechanism may be spun up
through repeated encounters in the BH cusp
\citep{2001ApJ...549..948A}, as well as during the final encounter
which ejects them (H.\ Perets, private communication).  Therefore, a
rotating BHB star may appear to be more like a B-star without detailed
spectroscopic evidence.  In fact, the HVS HE 0437-5439
\citep{2005Apj...634L.181E}, should not be ruled out as being a BHB
star based on its required spin value alone.

There is a curious link between the number of HVSs observed by
\citet{2006ApJ...647..303B} and the S-stars observed orbiting \sag.
Some S-stars may be the former companions to the HVSs from tidally
disrupted binaries \citep{2003ApJ...592..935G,2006MNRAS.368..221G,\perets}.
Interestingly, there exists a similar connection between HVSs
scattered off of the BHs and the S-stars as well, although perhaps not
one--to--one. For every strong encounter that produces a HVS, there is
likely another encounter which can bring the star closer to \sag,
perhaps kicking out the BH instead, similar to the scenario proposed
by \citet{2004ApJ...606L..21A}.  In this scenario, the scattering of a
small fraction of young stars from many previous disks into the inner
$0.04\,$pc may result in a population similar to the S-stars (O'Leary,
R.\ M.\ \& Loeb, A.\ 2007, in preparation).  However, we should note
that if the S-star are in fact the remnants of tidally disrupted
binaries, then this mechanism will only be secondary, as it is only
ejects a small fraction of such stars.

There is obviously considerable uncertainty in our models. The rate of
diffusion of stars both close to and far from \sag\ into eccentric
orbits is important to understanding both the mechanisms as well as
the source of the observed B-type HVSs in the Galactic halo. In both
cases, it is most likely that the stars formed on relatively circular
orbits, whether in a disk around \sag\ or in an inspiralling cluster.
Large scale simulations similar to those already done by
\citet{2006astro.ph..3280F} can help resolve the uncertainties of
relaxation, and with modifications, may be able to account for the
conditions of continuous star formation over long periods of time. For
low mass stars, physical collisions may deplete the total number of
stars, and reduce the rate.  In our discussion we have not considered
the effects of resonant relaxation
\citep{1996NewA....1..149R,2006ApJ...645.1152H} near \sag, which may
have two counteracting effects on our rate calculation.  Resonant
relaxation may flatten the cusp and deplete the number density of
stars and BHs in the innermost 0.01\,pc; at the same time, it may also
drive more massive stars into the same region producing more B-type
HVSs \citep{2006ApJ...645.1152H}.

In our analysis, we also neglected the migration of massive objects near
\sag.  If, as suggested by \citet{2006ApJ...641..319P}, many intermediate
mass BHs (IMBHs with masses $ \gtrsim 10^3\,\msun$) populate the inner pc
of the Galactic centre and merge with \sag\ every $\sim 10^7$--$10^8\,$yr,
then the cusp of stellar-mass BHs and stars would not regenerate fast
enough to produce HVSs through BH-star encounters
\citep{2006astro.ph..7455B} but instead could produce them through
IMBH-star encounters \citep{2005astro.ph..8193L}.  BHs with masses between
most stellar mass BHs and IMBHs (e.g. $\sim 50\,\msun$), may also form
through standard binary evolution \citep{2004ApJ...611.1068B}.  Such BHs
could considerably increase the rate of HVSs, even if they are relatively
rare.

\section*{Acknowledgements}

We would like to thank Reinhard Genzel for discussing his most recent
results on the stars near \sag, as well as Warren Brown and Hagai
Perets for their helpful discussions and comments on our manuscript. This
work was supported in part by Harvard University grants.


\end{document}